\documentclass[twocolumn,english]{article}
\usepackage[LGR,T1]{fontenc}
\usepackage[latin9]{inputenc}
\usepackage{float}
\usepackage{amssymb}
\usepackage{graphicx}
\usepackage{setspace}
\makeatletter

\DeclareRobustCommand{\greektext}{%
  \fontencoding{LGR}\selectfont\def\encodingdefault{LGR}}
\DeclareRobustCommand{\textgreek}[1]{\leavevmode{\greektext #1}}
\ProvideTextCommand{\~}{LGR}[1]{\char126#1}

\providecommand{\tabularnewline}{\\}

\newcommand{\lyxaddress}[1]{
\par {\raggedright #1
\vspace{1.4em}
\noindent\par}
}

\makeatother

\usepackage{babel}
\begin{document}

\title{Spectrum of Landau Levels in GaAs Quantum Wells}

\author{Imran Khan , Bipin Singh Koranga and Sunil Kumar}
\maketitle

\lyxaddress{Ramjas college (University of Delhi,) Delhi-110007, India}

\lyxaddress{Kirori Mal college (University of Delhi,) Delhi-110007, India}

\lyxaddress{Ramjas college (University of Delhi,) Delhi-110007, India}
\begin{abstract}
We have studied the electroluminescence spectra from an n-i-p LED
as a function of magnetic field. This sample incorporated three GaAs
quantum wells in the intrinsic region. This device had excess n-type
doping and as a result. The quantum wells were populated by a 2D Landua
electron gas. The broad B=0 field emission band evolved into a series
of discrete features in the presence of a magnetic field. These were
identified as inter-band transitions between the \ensuremath{\ell}
= 0, 1, and 2: Landau levels associated with the $e_{1}$ and $h_{1}$
sub-bands, with the selection rule \textgreek{D}\ensuremath{\ell}
= 0. The EL spectra were analyzed in their \ensuremath{\sigma}+ (LCP)
and \ensuremath{\sigma}- (RCP) components. An energy splitting between
the two polarized components was observed for each Landau level transition.
This was found to be equal to the sum of the conduction and valence
band spin splittings. We used the know value of elctron\textquoteright s
g-factor {[}11{]} to determine the valence band spin splittings. Our
experimental values were compared to the numerically calculated values
shown in {[}1{]} and were found to be in reasonable agreement.
\end{abstract}

\section{Introduction}

The injection of spin-polarized electrons from a ferromagnetic Fe
contact into empty quantum wells has been studied in detail {[}1{]}
\& {[}2{]}. We have also studied injection of electrons into GaAs
quantum wells which are occupied by a dilute electron gas (areal electron
density less than $10^{11}cm^{-2}$ . In these devices the emission
is excitonic in nature. In this chapter we describe magneto-EL studies
of Spin-LEDs which incorporate quantum wells occupied by a dense electron
gas, areal electron density less than $10^{12}cm^{-2}$. In this system
the excitons are screened {[}3{]}, {[}4{]} \& {[}5{]} and only interband
transitions among the $e_{1}$ and $h_{1}$ subband Landau levels
are observed in the emission spectra\textbf{.Addd\textemdash -see}

\section{Bandstructure of Sample}

In table 1 we give a description of the sample used in our studies.

\begin{table}[p]
\begin{tabular}{|c|l|c|}
\hline 
Layer & Description  & Thickness (A)\tabularnewline
\hline 
\hline 
buffer & p+ GaAs & 1250\tabularnewline
\hline 
p+ barrier & p+ AlGaAs (25\% Al) & 250\tabularnewline
\hline 
Undoped barrier & AlGaAs & 250\tabularnewline
\hline 
Quantum Well \#1 & GaAs & 125\tabularnewline
\hline 
Undoped barrier & AlGaAs (25\% Al) & 300\tabularnewline
\hline 
Quantum Well \#2 & GaAs & 125\tabularnewline
\hline 
Undoped barrier & AlGaAs (25\% Al & 100\tabularnewline
\hline 
Quantum Well \#3  & GaAs & 125 \tabularnewline
\hline 
Undoped barrier & AlGaAs (25\% Al) & 100 \tabularnewline
\hline 
n \textasciitilde{} 1e17 & n- AlGaAs & 430 \tabularnewline
\hline 
transistion &  transition AlGaAs & 150 \tabularnewline
\hline 
n \textasciitilde{} 1e19 & n+ AlGaAs & 150 \tabularnewline
\hline 
cap &  Fe & 160\tabularnewline
\hline 
\end{tabular}

\caption{Description of the sample used and its dimension}
\end{table}

A schematic diagram of the bandstructure of the device studied is
shown in fig.(1) under flat band and high forward bias conditions.
It consists of three 125 $A^{o}$ GaAs quantum wells separated by
300 $A^{o}$ of AlGaAs barriers.

\begin{figure}[H]
\includegraphics[scale=0.35]{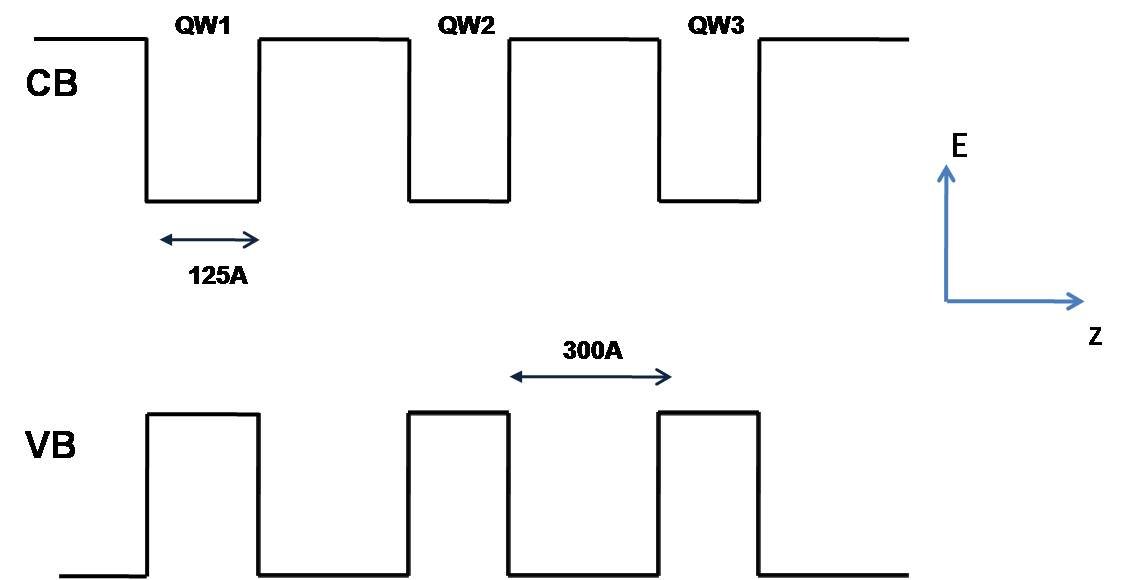}

\caption{Schematic diagram of the hetrostructure studied.}
\end{figure}

The doping level of the n-type section of the n-i-p junction was higher
than the doping of its p-type component. The calculated band diagram
at zero bias is shown in fig.(2). The diagram was generated by \textquotedblleft 1D
Poisson\textquotedblright{} program, which is a program for calculating
energy band diagrams for semiconductor structures written by Greg
Snider from University of Notre Dame {[}ref{]} (www.nd.edu/\textasciitilde{}gsnider/
). We note that the $e_{1}$ subband of all three quantum wells in
this LED lie above the Fermi level( dotted line) ; as a result at
V = 0 all three quantum wells are empty.

\begin{figure}[H]
\includegraphics[scale=0.25]{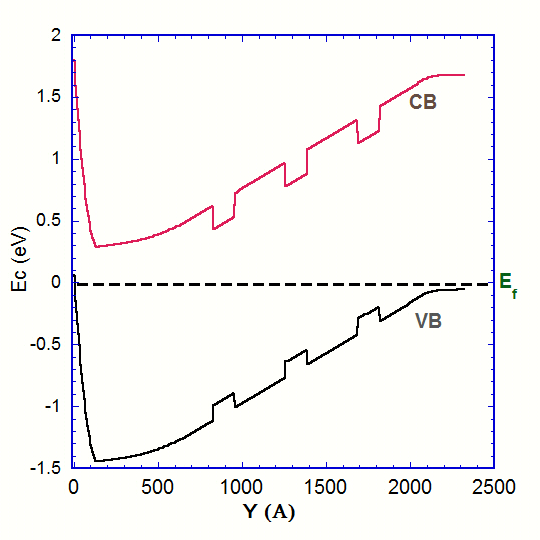}

\caption{Calculated band diagram at zero bias{[}ref{]}.}
\end{figure}

\section{Results and Discussion:}

The two circularly polarized components (\textgreek{sv}+/red and \textgreek{sv}-/black
) of the EL spectra recorded at T = 7 K in the presence of a magnetic
field B = 5 tesla under low bias conditions is shown in fig.(3). Under
these conditions the excess donors in the structure have not released
their electrons into the quantum wells and thus the quantum wells
are empty. As a result, the emission feature at 12400 $cm^{-1}$ of
fig.(3) is excitonic in nature ($e_{1}\,h_{1}$ exciton).

\begin{figure}[H]
\includegraphics[scale=0.25]{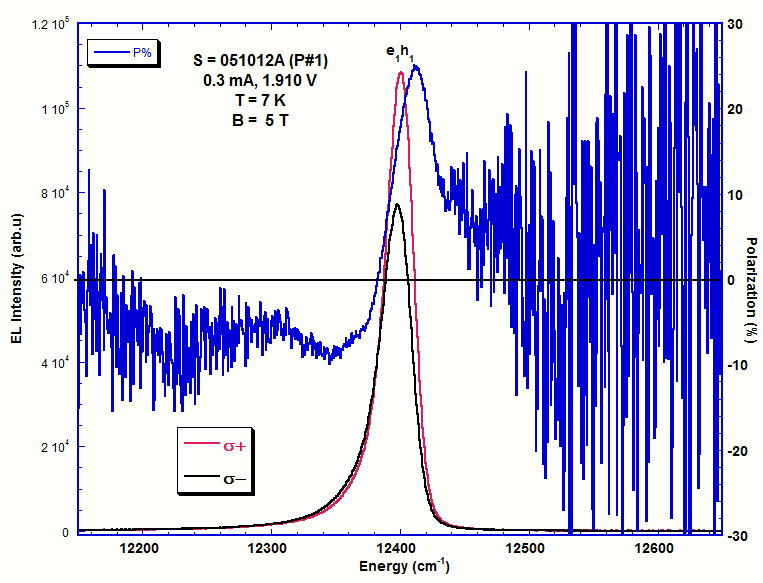}

\caption{EL spectra at T = 7 K and B = 5 T..}
\end{figure}

The blue line in fig.(3) represents the circular polarization P of
the spectrum. We note that the polarization maximum is on the high
energy side of the EL peak but still lies within the linewidth of
the emission. We attribute the maximum in P at 12411 $cm^{-1}$ to
the free exciton, and the maximum in EL intensity at 12400 $cm^{-1}$
to the bound excitons. In fig.(4) we plot the circular polarization
P versus magnetic field B at the polarization maximum (red circles)
and at the EL intensity maximum (black squares).

\begin{figure}
\includegraphics[scale=0.25]{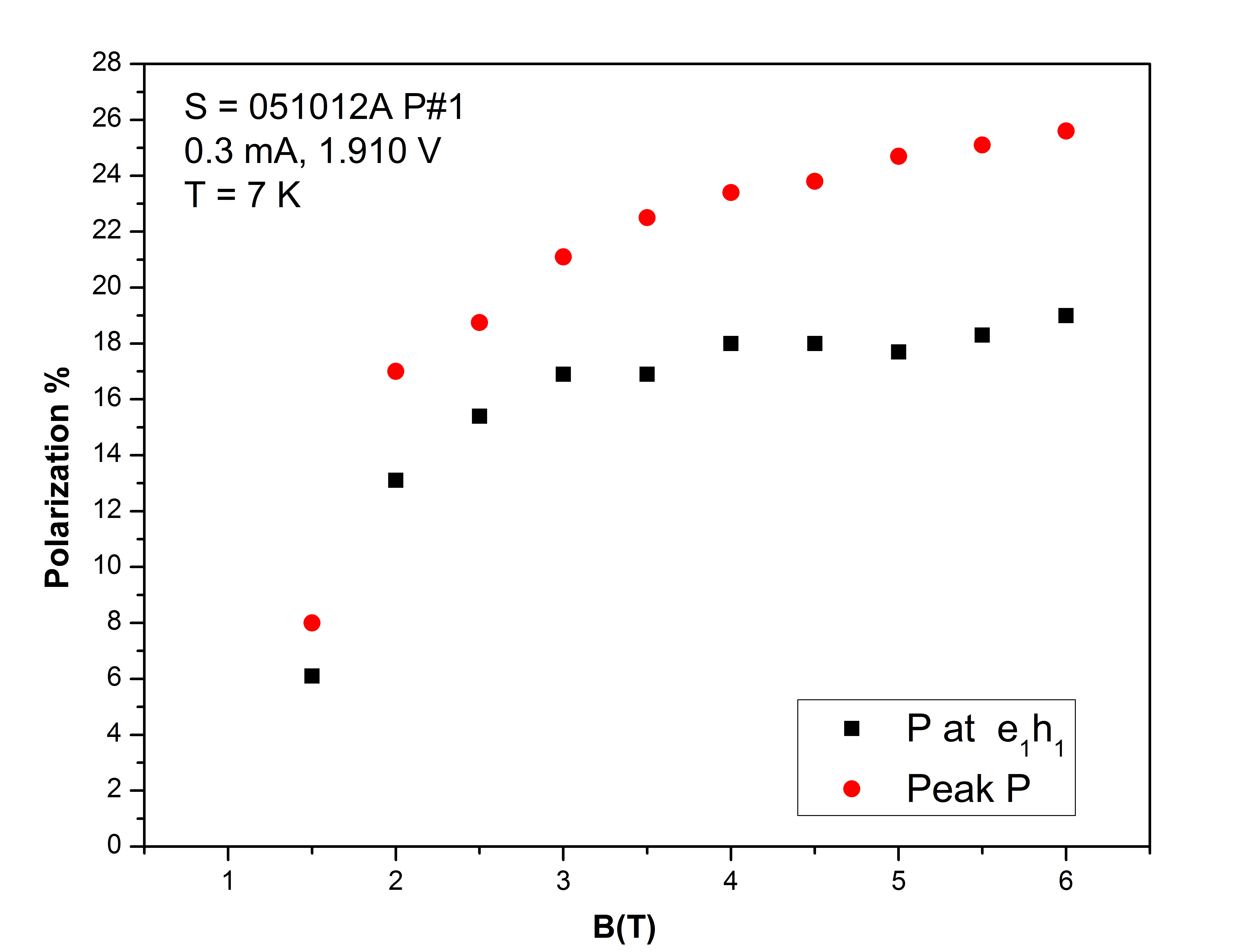}

\caption{Circular polarization P versus magnetic field B.}
\end{figure}

Both sets of data show clear evidence of spin injection from the top
Fe layer. The polarization mirrors the out of plane magnetization
of the Fe contact{[}2{]} The bound excitons (black squares) show polarizations
below that of the free exciton (red circles) because in addition to
the recombining electron-hole pair, the impurity atom on which the
exciton is bound is involved. The data of fig.(4) clearly show that
the Fe/AlGaAs(n) Schottky barrier is an efficient injector of spin
polarized electrons into the empty quantum wells. The device current
was increased to obtain the flat band conditions shown in fig.(1).
The expectation was that the excess donors would release their electrons
into the three quantum wells and that these electrons would be equallydistributed
among the three quantum wells. The zero field EL spectrum at T = 7
K, for a current of 1 mA is shown in fig.(5)

\begin{figure}[H]
\includegraphics[scale=0.25]{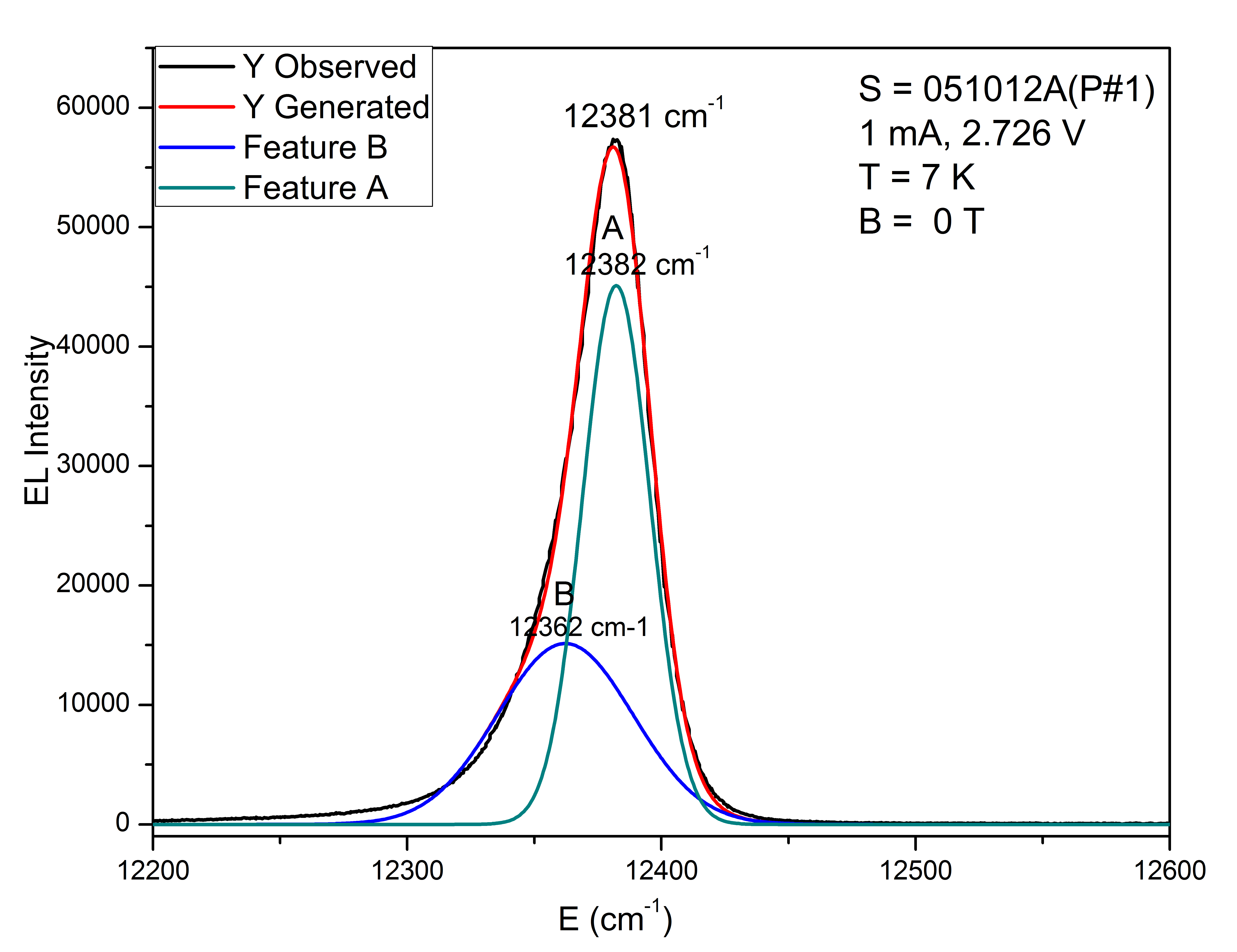}

\caption{Zero field EL spectrum at T = 6 K, for a current of 1 mA.}
\end{figure}

The spectrum contains two unresolved features which can be separated
using the line fitting program \textquotedblleft Peakfit\textquotedblright .
The calculated features are labeled \textquotedblleft A\textquotedblright{}
and \textquotedblleft B\textquotedblright{} and are also plotted in
fig.(5). Feature A (green line) at 12382 $cm^{-1}$ is identified
as the $e_{1}\,\,h_{1}$ exciton, while feature B (blue line) at 12362$cm^{-1}$
is attributed to the electron-recombination of the two-dimensional
electron gas with holes {[}6{]},{[}7{]}\&{[}8{]} . It is clear from
the EL spectrum of fig.(5) that the electrons have populated one or
two quantum wells; feature B is associated with the populated well(s).
At least one quantum well is empty; feature A is the excitonic emission
from one or more empty wells. In order to verify this, we applieda
magnetic field perpendicular to the GaAs layers. The EL spectrum analyzed
as \textgreek{sv}- is shown in fig.(6) for B = 6 tesla.

\begin{figure}[H]
\includegraphics[scale=0.25]{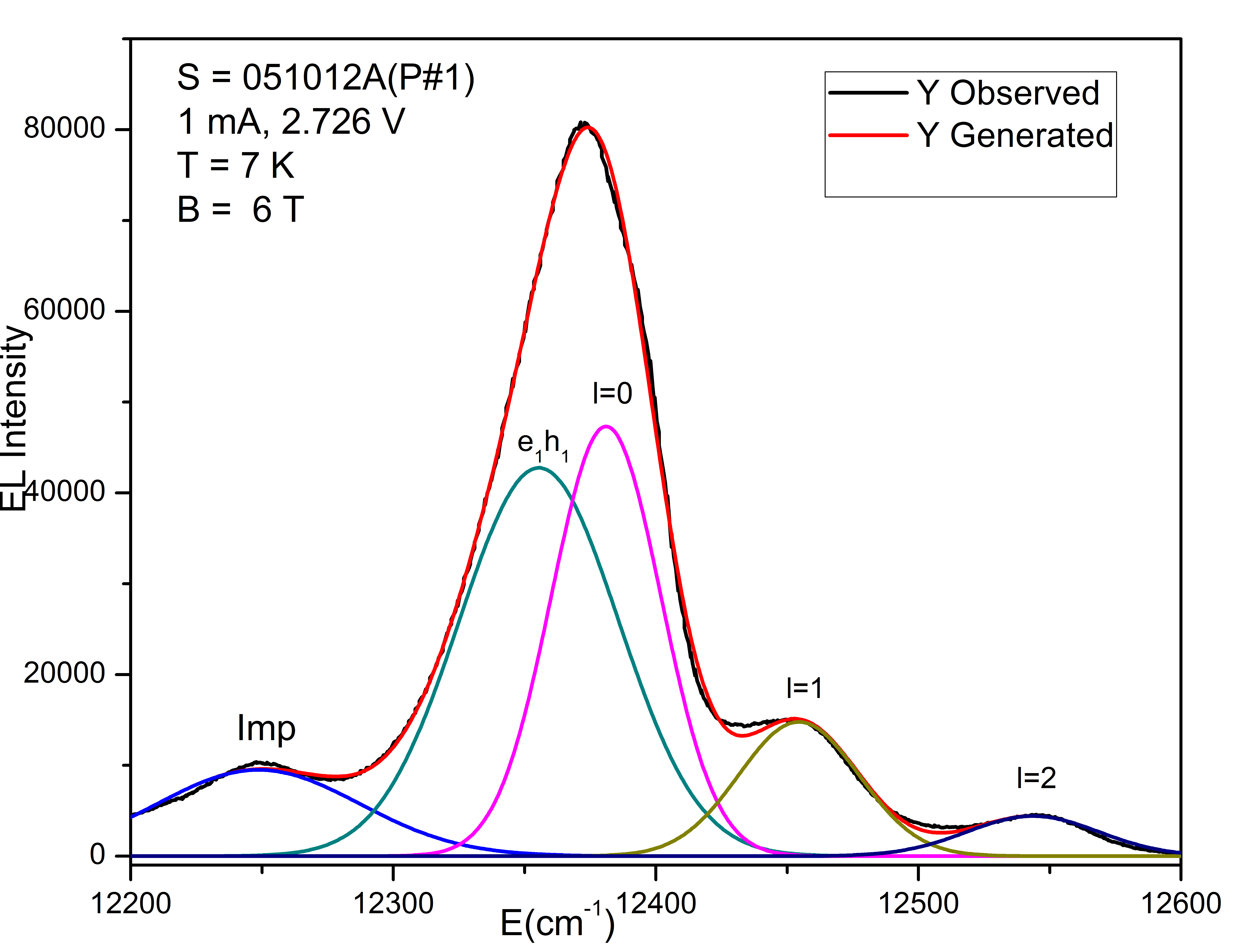}

\caption{The EL spectrum analyzed as \textgreek{sv}- for B = 6 tesla..}
\end{figure}

On the same figure we also give the various spectral components into
which the EL was deconvoluted using the \textquotedblleft Peakfit\textquotedblright{}
program. The feature labeled \textquotedblleft Imp\textquotedblright{}
at 12249 $cm^{-1}$ becomes weaker as the device temperature is raised
and then totally disappears. It is therefore attributed to an impurity-related
transition. Feature labeled \textquotedblleft $e_{1\,\,}h_{1}$ \textquotedblright{}
at 12373 $cm^{-1}$ is identified as the unscreened exciton from an
empty quantum well. The features labeled 0 at 12380 $cm^{-1}$ , 1
at 12454 $cm^{-1}$ , and 2 at 12544 $cm^{-1}$ are attributed to
interband transitions between the 0, 1, and 2 conduction and valence
band Landau transitions, respectively associated with feature B in
fig.(5) Thus at low temperatures,we have two types of quantum wells
in this device. One quantum well is empty and is associated with the
$e_{1}\,h_{1}$ exciton and two quantum wells that are occupied by
a two- dimensional electron gas. Under some bias conditions we actually
see two sets of Landau levels which are slightly displaced in energy.
That indicates that the remaining two quantum wells are occupied by
electrons but the two electron populations are unequal, which accounts
for the different bandgap renormalization {[}9{]} \& {[}10{]}. Under
these conditions we found it impossible to disentangle the polarizations
of the two types of quantum wells. The study of spin injection will
have to be carried out on pairs of Fe Spin-LEDs: one LED in which
the quantum well is empty (equal n- and p-type doping) and the other
in which the quantum well is occupied by a two-dimensional electron
gas (excess n-type doping).The comparison of the polarizations between
the two emission spectra is expected to give us a clear idea of what
happens when spin polarized electrons are injected in a quantum well
which is already occupied by a dense electron gas. Given these constraints
we decided to raise the sample temperature in the hope thatthis will
result in populating the three quantum wells with equal number of
electrons. field The zero emission spectrum at T = 75 K for I = 1.5
mA is shown in fig.(7). It closely resembles the emission spectra
from n-type modulation doped QWs.

\begin{figure}[H]
\includegraphics[scale=0.25]{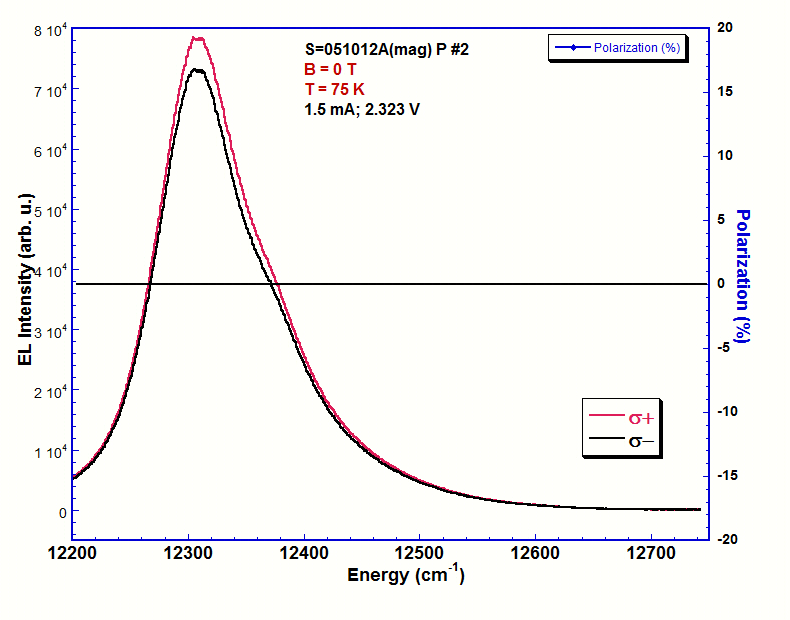}

\caption{Zero field emission spectrum at T = 75 K for I = 1.5 mA.}
\end{figure}

In the presence of an external magnetic field applied perpendicular
to the heterostructure\textquoteright s layers the continuum breaks
into a Landau fan and there is no evidence of the $e_{1}\,h_{1}$
exciton. In the presence of a magnetic field we have the appearance
of distinct features in the EL spectra. An example is given in fig
4.7 for B = 6.5 T.

\begin{figure}[H]
\includegraphics[scale=0.25]{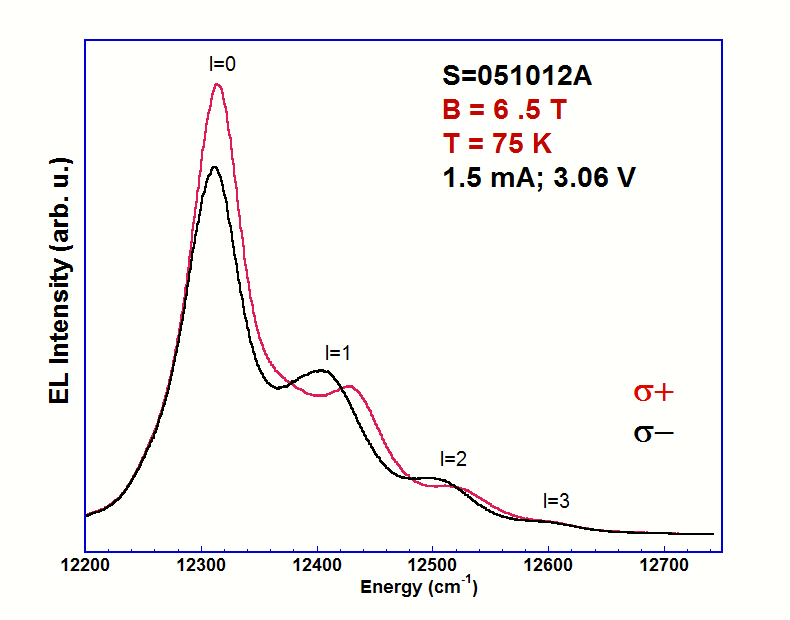}

\caption{EL Spectra in presence of a magnetic field of B=6.5T.}
\end{figure}

Both circular polarization components (\textgreek{sv}+ and \textgreek{sv}-
) are shown in the same figure. The emission features are labeled
using the quantum number of the Landau levels associated with the
$e_{1}$ and $h_{1}$ subbands that are involved in a particular EL
feature. A schematic diagram of the energy states in the conduction
and valence band and the allowed interband transitions are shown in
fig.(9).

\begin{figure}[H]
\includegraphics[scale=0.25]{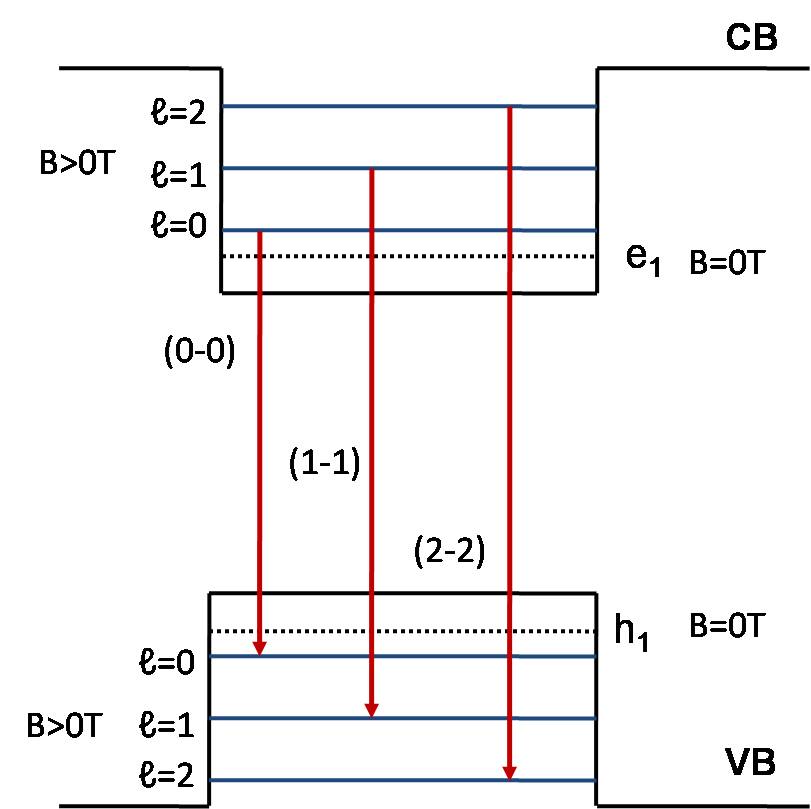}

\caption{Schematic diagram of the energy states in the conduction and valence
band.}
\end{figure}

The interband transitions occur between Landau levels with the same
quantum number i.e. they obey the selection rule 0 . In this diagram
the spin splitting has been omitted. The various inter-band transitions
among the spin-split Landua levels are shown in fig.(11). In fig.(10)
we plot the energies of the inter-band transitions measured in our
experiment as function of magnetic field.

\begin{figure}
\includegraphics[scale=0.25]{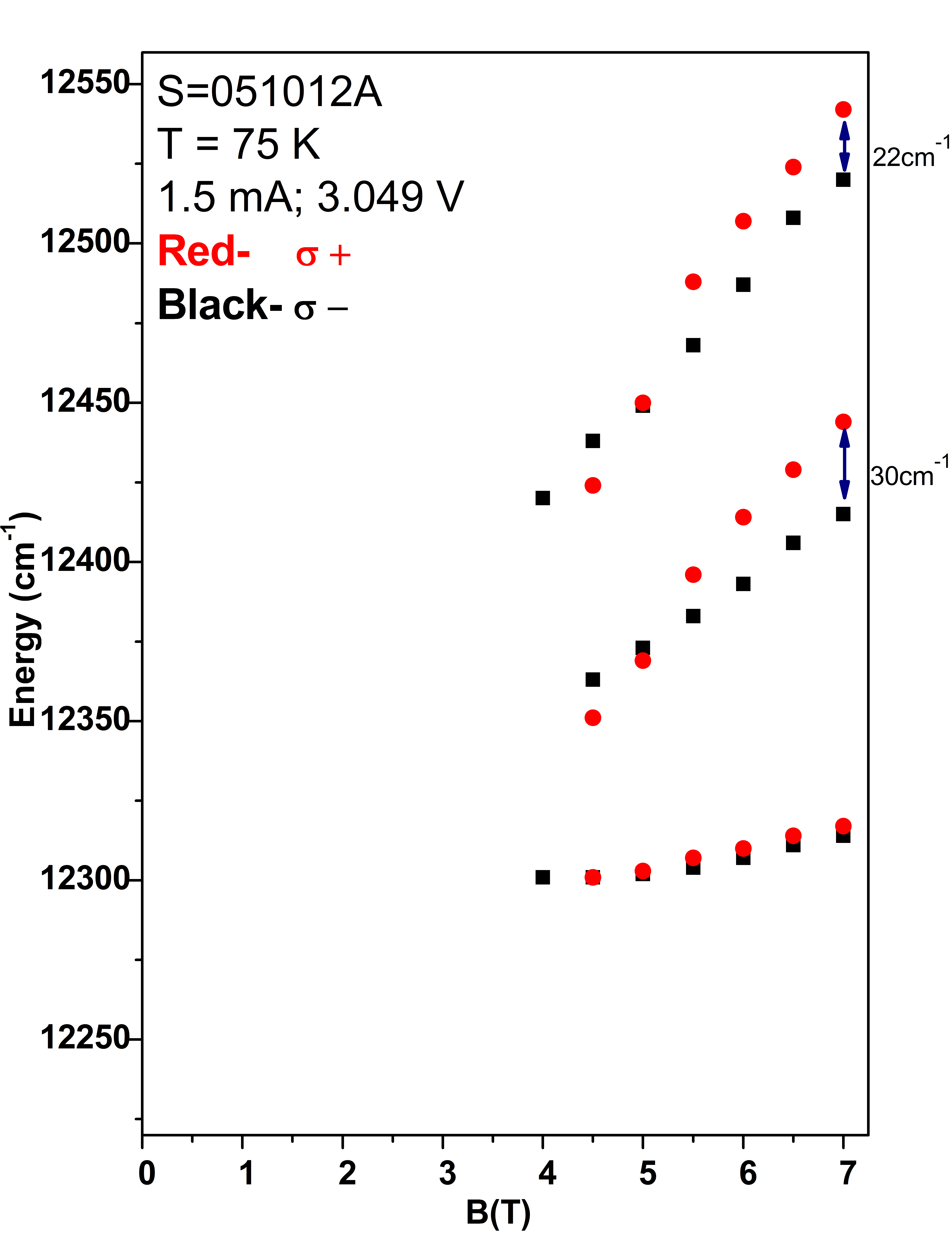}

\caption{Energies of the interband transitions as function of magnetic field..}
\end{figure}

We have identified EL features associated with the 0 , 1, and 2 Landau
levels of the $e_{1}$ and $h_{1}$ confinement sub-bands. Each EL
feature is analyzed in its andcomponents, and the energy of each component
is plotted separately. A schematic diagram of the conduction and valence
sub-band Landau level spin splitting, as well as the allowed transitions
in the Faraday geometry are shown in fig.(11). The spin splittings
have been greatly exaggerated for the sake of clarity.

\begin{figure}[H]
\includegraphics[scale=0.25]{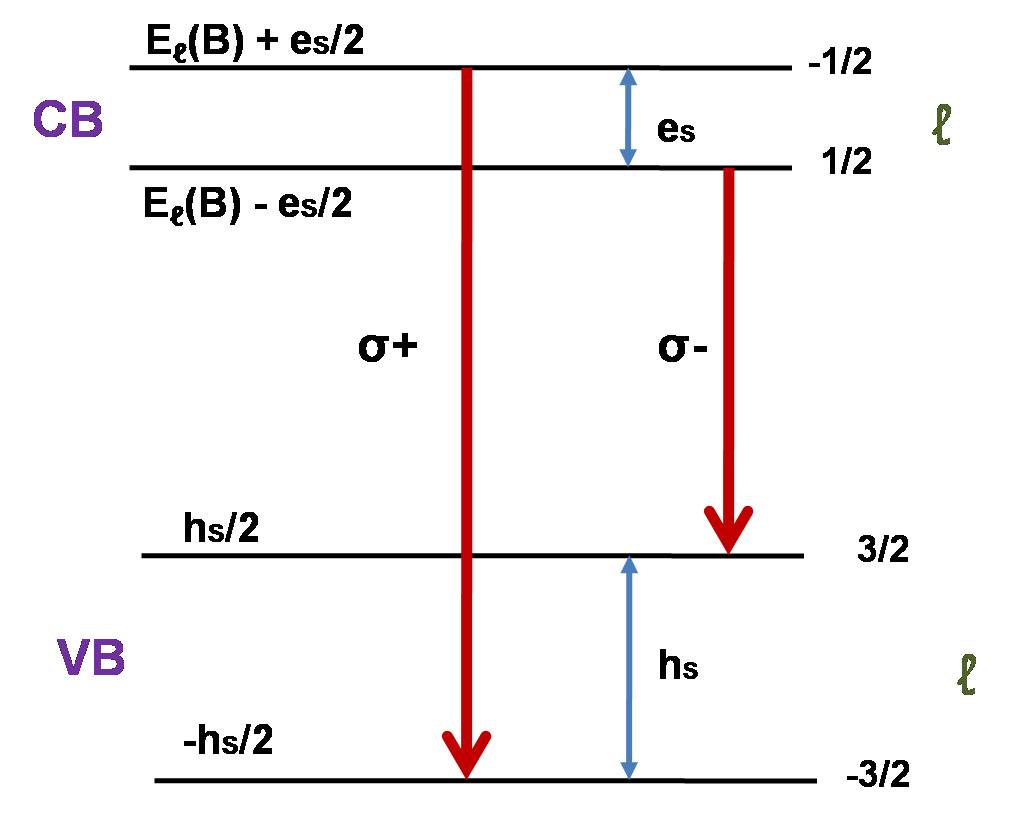}

\caption{Schematic diagram of the conduction and valence subbands Landau level
spin splitting..}
\end{figure}

The energies of the and components of the photon associated with the
recombination among the Landau level with quantum number are given
by the equations:

\begin{equation}
E_{l}(\sigma+)=E_{l}(B)+\frac{e_{s}+h_{s}}{2},
\end{equation}

\begin{equation}
E_{l}(\sigma-)=E_{l}(B)-\frac{e_{s}+h_{s}}{2},
\end{equation}

here

\[
E_{l}(B)=E_{g}^{*}+\left(l+\frac{1}{2}\right)\left(\hbar\omega_{ce}+\hbar\omega_{ch}\right)
\]

$e_{s}$ and $h_{s}$ are the electron and hole spin splittings, respectively;
ce and ch are the electron and hole cyclotron energies, respectively.
The splittings between the $E_{l}(\sigma+)$ and $E_{l}(\sigma-)$
are equal to $e_{s}h_{s}$. The conduction band spin splitting $e_{s}$
is equal to $g^{*}\mu_{B}B$ and can be calculated because the effective
Landé g factor for GaAs conduction band has been measured {[}11{]}.
From our experimental values of the difference $E_{l}(\sigma+)$ -$E_{l}(\sigma-)$,we
can extract the hole spin splitting and compare it with the calculated
values {[}12{]} shown in fig.(12).

\begin{figure}[H]
\includegraphics[scale=0.25]{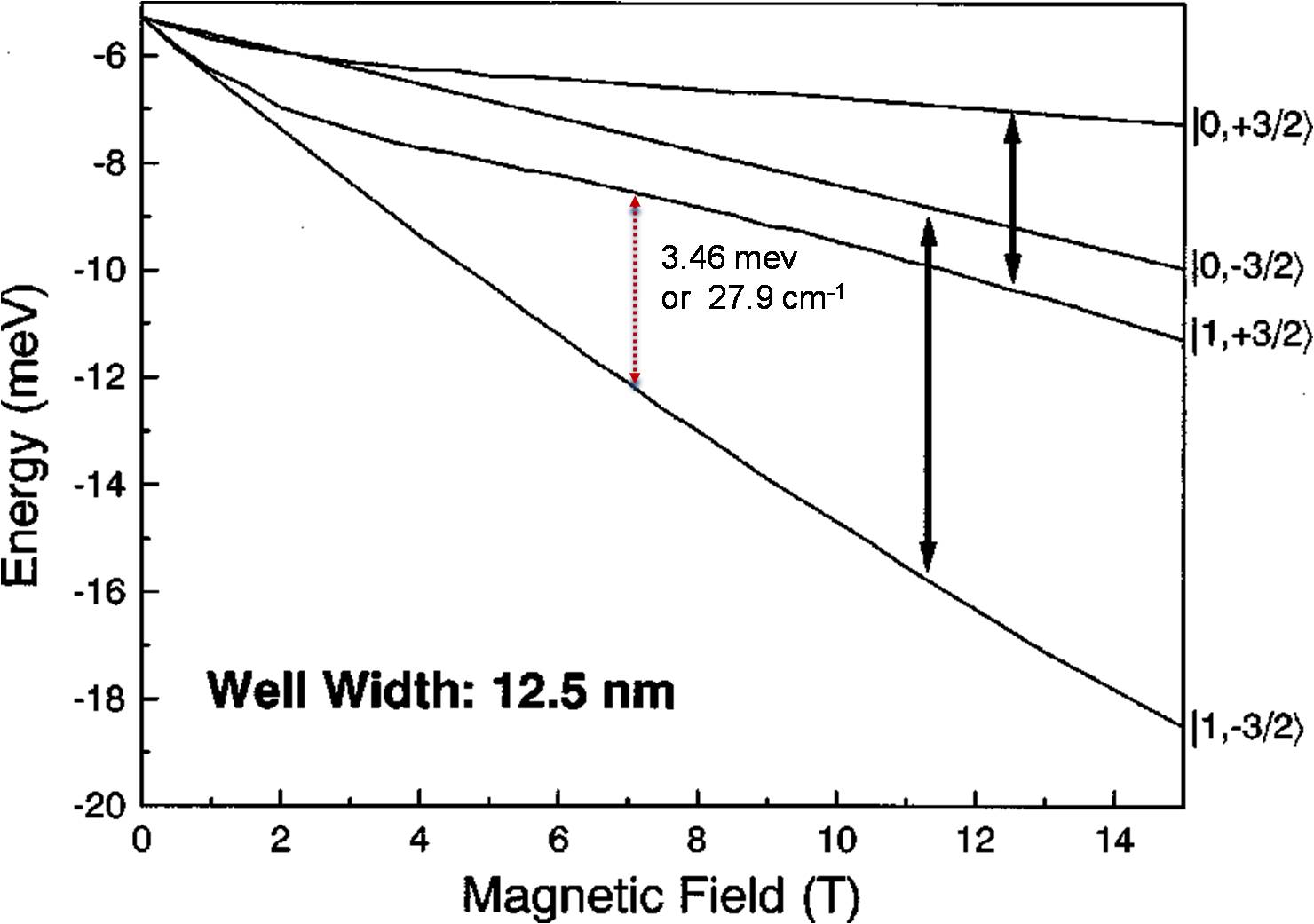}

\caption{Calculated Landau-level structure for holes in the valence band of
a GaAs/Al0.3Ga0.7As quantum well structure with 12.5-nm wells. 1.}
\end{figure}

Earlier study {[}12{]} has measured the hole cyclotron energies $\hbar\omega_{ch}$
for the $m_{j}=\pm\frac{3}{2}$ holes involved in the transitions
{[}12{]}. The cyclotron energies are indicated by the double headed
arrows in fig.(12). In our experiment we have measured the energy
indicated by the dotted double arrow in fig.(12). transition is too
small to be determined reliably. The splitting for the 0 Landau for
1 our experimental value for $e_{s}h_{s}$= 29.8 $cm^{-1}$ at B =
7 tesla. The calculated value for $e_{s}$ is 1.41 $cm^{-1}$ from
which we extract a value of 28.4 $cm^{-1}$ for $h_{s}$ . The calculated
value is 27.9 $cm^{-1}$ at B = 7 tesla. For 2 we get an experimental
splitting of 22 $cm^{-1}$ . Unfortunately we do not have calculated
energies for the spin components of this Landau level.

\section{Conclusions\textemdash add-see}

The work described in this chapter has shown that magneto-EL studies
of quantum wells is a useful tool for the study of the energy states
of the system. In the past there have been extensive magneto-PL studies
of quantum wells and the concern in these experiment is to minimize
the excess energy introduced into the system from the fact that the
existing photon energy has to be above the inter-band transition under
study. In EL experiments the electrons and holes are introduced in
the well without an excess energy. We plan to carry out magneto EL
studies in locally grown LEDs. In addition we plan to pursue studies
of spin LEDs which incorporate only one quantum well. In this these
experiments we intend to measure the polarization emission characteristics
from a series of diodes in which the quantum well have the same well
width but have different electron concentration.

\end{document}